\documentclass[aps,prb,reprint,superscriptaddress, longbibliography]{revtex4-1}
\usepackage{H1}
\usepackage{amsmath}
\usepackage[pdftex,colorlinks=true]{hyperref}
\hypersetup{citecolor = blue,urlcolor=blue}
\usepackage[normalem]{ulem}
\usepackage{amssymb}
\usepackage{amsthm}
\usepackage{amsfonts}
\usepackage{bbm}
\usepackage{array}
\usepackage{braket}
\usepackage[dvipsnames]{xcolor}

\newcommand{\Sztot}{S^z_{\rm tot}}
\newcommand{\Sxtot}{S^x_{\rm tot}}
\newcommand{\X}{S^x}
\newcommand{\Y}{S^y}
\newcommand{\Z}{S^z}

\newcommand{\Heff}{H_{\rm{eff}}}
\newcommand{\Ho}{\hat{\mathcal{H}}_F^{(0)}}

\newcommand{\Tr}{ \mbox{Tr}}

\begin{document}
\title{Prethermalization without temperature}

\author{David J. Luitz}
\affiliation{Max-Planck-Institut f\"{u}r Physik komplexer Systeme, 01187 Dresden, Germany}

\author{Roderich Moessner}
\affiliation{Max-Planck-Institut f\"{u}r Physik komplexer Systeme, 01187 Dresden, Germany}

\author{S. L. Sondhi}
\affiliation{Department of Physics, Princeton University, Princeton, NJ 08544}

\author{Vedika Khemani}
\affiliation{Department of Physics, Harvard University, Cambridge, MA 02138, USA}
\affiliation{Department of Physics, Stanford University, Stanford, CA 94305, USA}

\date\today

\begin{abstract}
While a clean driven system generically absorbs energy until it reaches `infinite temperature', it may do so very slowly exhibiting what is known as a prethermal regime. Here, we show that the emergence of an additional approximately conserved quantity in a periodically driven (Floquet) system can give rise to an analogous long-lived regime. This can allow for non-trivial dynamics, even from initial states that are at a high or infinite temperature with respect to an effective Hamiltonian governing the prethermal dynamics. 
We present concrete settings with such a prethermal regime, one with a period-doubled (time-crystalline) response. We also present a direct diagnostic to distinguish this prethermal phenomenon from its infinitely long-lived many-body localised cousin. 
We apply these insights to a model of the recent NMR experiments by 
 Rovny \textit{et al.}, [\href{https://doi.org/10.1103/PhysRevLett.120.180603}{Phys. Rev. Lett. \textbf{120}, 180603 (2018)}] which, intriguingly, detected signatures of a Floquet time crystal in a clean three-dimensional material. We show that a mild but subtle variation of their driving protocol can increase the lifetime of the time-crystalline signal by orders of magnitude.
\end{abstract}

\maketitle

\section{Introduction} The study of quantum systems out of equilibrium has led to the identification of fundamentally new phenomena, such as the discrete time crystal (DTC) in periodically driven (Floquet) systems~\cite{khemani_phase_2016,else_floquet_2016, CVS, MoessnerSondhiReview, briefhistory, ElseTCReview, SachaReview}. In a generic many-body system, periodic driving leads to heating to a featureless `infinite temperature' state, appropriate to maximizing entropy in a system with no conservation laws~\cite{RigolPeriodicHeating,LazaridesHeating,Ponte15b}. The only known generic mechanism for avoiding this heating~\cite{Lazarides14,Ponte15,Ponte15b,Abanin14, RigolPeriodicHeating}---in the asymptotic limit of large systems and late times---relies on the phenomenon of many-body localization (MBL) in disordered, interacting systems~\cite{Anderson58, Basko06, gornyi_interacting_2005, PalHuse, ProsenMBL2008, OganesyanHuse, Luitz15, Imbrie2016, Nandkishore14, EhudMBLRMP}. This permits the existence of non-trivial MBL Floquet phases~\cite{khemani_phase_2016}, the DTC being a paradigmatic example which displays a novel form of long-range \emph{spatiotemporal} order---breaking both the discrete time-translation symmetry of the periodic drive \emph{and} an emergent (spatial) Ising symmetry~\cite{khemani_phase_2016, CVS, TTSBRep, briefhistory}.

Many-body localization requires a number of idealized conditions (for example, perfect environmental isolation and short ranged interactions) that may not always be realized in a given experimental set up. Nevertheless, even absent MBL, it was shown that the heating time can be made exponentially large in some dimensionless system parameters, $t_h \sim O(\exp(\omega/J))$, when the driving frequency, $\omega$, is large compared to the local energy scales in the system, $\sim J$~\cite{DimaPrethermal_linearresponse, DimaPrethermal_Heff, DimaPrethermal_rigorous, MoriPrethermal1, MoriPrethermal2}. Intuitively, absorbing one `quantum' of energy $\omega$ from the drive requires the rearrangement of many local degrees of freedom with energy scales $J$, which is a high-order process leading to parametrically slow heating. 

In the ``prethermal" regime prior to heating, $t < t_h$, the system can display non-trivial dynamics and is well described by a (quasi-local) time-independent ``effective Hamiltonian" $H_{\rm eff}$ that captures the dynamics of the system out to an exponentially long time~\cite{DimaPrethermal_rigorous,DimaPrethermal_Heff,MoriPrethermal1, MoriPrethermal2}. Building on this, it was shown in Ref~\cite{else_prethermal_2017} that a DTC can be realized for an extended prethermal regime, even absent MBL, if one arranges for (a slightly generalized) $H_{\rm eff}$ to additionally display an emergent Ising symmetry, with a spontaneous Ising symmetry breaking transition at some temperature $T_c$. Then, upon starting from a symmetry-broken initial state at a low temperature below $T_c$, the system can display oscillations of the Ising order parameter at twice the driving period. At late times, the system eventually heats to infinite temperature and $\Heff$ ceases to be a good description. We will refer to such prethermal time-crystals that rely on spontaneous symmetry breaking (SSB) as ``prethermal SSB DTCs".  

Intriguingly, a recent NMR experiment on a \emph{clean}, periodically driven three dimensional crystal observed signatures of time-crystallinity, despite being far from any MBL regime~\cite{rovny_observation_2018, RovnyPRB}. The experiment measured the global magnetization of the sample, and observed period doubled oscillations for the duration of the experimental coherence time (about a 100 driving periods). Despite the almost complete lack of disorder, the observed signal was very similar to that observed in two earlier experiments, on disordered nitrogen vacancy centers~\cite{MishaTCExp} and trapped ions~\cite{MonroeTCExp}, that were closer in spirit to MBL TCs due to slow disorder-impeded thermalization~\cite{CriticalTCPRL}. 

A natural conjecture is that the clean NMR experiment may be seeing a prethermal SSB DTC \`{a} la Ref.~\cite{else_prethermal_2017}. However, the experiment prepares a  weakly polarized initial state that is at an extremely \emph{high} temperature (vastly in excess of the strength of the dipolar interactions in the crystal). This does not satisfy the requirement in ~\cite{else_prethermal_2017} for starting with a symmetry-broken initial state at a \emph{low}-temperature with respect to $\Heff$. 

Thus, the NMR results do not fit into any existing framework of Floquet MBL (or prethermal) order, and call for a new theory. We identify \emph{the emergence of a long-lived approximately conserved quantity} as the crucial missing ingredient. 
The existence of this conserved quantity stabilizes the time crystalline behaviour and provides a prethermal window via a long timescale on which this conservation law is eventually destroyed. This conservation law may or may not be accompanied by the presence of approximate long-lived energy conservation (\emph{i.e.} the existence of a local time-independent $\Heff$) in previously identified prethermal phenomena, thereby extending these qualitatively. We also emphasize that the existence of this conservation law does not, in turn, require any (conventional) spontaneous symmetry breaking.

To make contact with the experiment, we arrange for the emergence of a long-lived $U(1)$ symmetry, that is approximately the total spin (or global magnetization $M$) along the $z$ direction. We primarily focus on cases where there is also long-lived energy conservation and hence an $\Heff$. Here one can show that dynamics from initial states at \emph{infinite} temperature but non-zero magnetization density can nevertheless show non-trivial dynamics (such as long-lived oscillations of $M(t)$) for a long period of time, thereby allowing for the apparently oxymoronic notion of \emph{prethermalization without temperature}.

One of the insights deriving from our analysis is that a prethermal DTC signal is most stable for parameter values which may not have been a priori obvious. In particular, a well-known route to realizing an approximate $U(1)$ symmetry in a time-independent system is to apply a large magnetic field in the, say, $z$-direction~\cite{DimaPrethermal_rigorous}. However, as we discuss below, this is not as straightforward in some natural Floquet settings since the stroboscopic nature of the Floquet unitary does not allow for the accumulation of arbitrarily large phases. 

The crispest mechanism for realizing the physics we have in mind entails engineering the desired emergent symmetry to leading order in $H_{\rm eff}$, with residual symmetry breaking perturbations arising only at higher orders in a small paramater $\epsilon/\omega$. In more detail, a fundamental object of interest in a Floquet system is the time-evolution operator over one driving period $T$, defined as $U(T) = \mathcal{T} e^{-i \int_0^t dt H(t)}$. This can be used to formally define a (non-unique) `Floquet Hamiltonian' $\hat{H}_F$ via $\hat{U}(T) = \mathcal{T} e^{-i\int_0^T {\rm d}t\, \hat{H}(t)/\hbar} \equiv e^{-i \hat{H}_F T/\hbar}$, where the operator $\hat{H}_F$ is generally highly non-local in a many-body system. When $\omega$ is large compared to the local energy scales of the problem, one can perform a high-frequency asymptotic expansion for $\hat{H}_F$ in powers of $1/\omega$, $\hat{H}_F = \sum_n (1/\omega)^n \, \hat{\mathcal{H}}_F^{(n)}$; the leading-order term $\hat{\mathcal H}_F^{(0)}$ is the time-averaged Hamiltonian, while higher-order terms are progressively longer-ranged   and contribute significantly to the dynamics only at correspondingly later times. While ultimately divergent, this expansion looks convergent out to some optimal order $n_\text{opt} = O(\omega/J)$. Truncating the expansion at this order yields $H_{\rm eff}$ which is an exponentially accurate approximation to the Floquet time evolution $\hat{U}(T)$, thereby setting the rate of heating to be exponentially small~\cite{DimaPrethermal_rigorous,DimaPrethermal_Heff,MoriPrethermal1, MoriPrethermal2}. If $\hat{\mathcal{H}}_F^{(0)}$ has the desired symmetry, with violations coming in at higher orders with strength $(\epsilon/\omega)^n$ (where $\epsilon$ is an independently chosen small parameter), then the time-scale on which the symmetry is destroyed can be made parametrically large for small $\epsilon$ and large $\omega$.

In sum, our work (i) widens the scope  of Floquet prethermalization, (ii) expands the toolkit for using Floquet system to generate dynamics with novel drives and symmetries and (iii) sheds light on the mystery of the NMR time-crystal experiment. In particular, we also predict that a slight and straightforward modification of the original experimental NMR protocol \cite{rovny_observation_2018, RovnyPRB} --{a judicious choice of an optimal magnetic field driving protocol}-- will \textit{exponentially} enhance the many-body lifetime of the observed DTC. 

The rest of this manuscript is structured as follows. In Section~\ref{sec:model}, we present the drive studied in the NMR DTC experiment, and introduce a family of short-range interacting one-dimensional spin 1/2 models inspired by the experiment as model systems to provide evidence confirming our picture. We then analyse the regimes of thermalization for our model drives in Section~\ref{sec:heating}, showing how one can engineer a long-lived approximate emergent $U(1)$ conservation that can show non-trivial magnetization dynamics even at infinite temperature and enhance the lifetime of the DTC signal observed in the NMR experiment. Section~\ref{sec:distinguish} provides concrete signatures for distinguishing between the different MBL and prethermal regimes in experiment, while Section~\ref{sec:conclusion} concludes with a summary and outlook.

\section{NMR Floquet drive} 
\label{sec:model}

In order to keep this work self-contained, we briefly summarize the pertinent details of the NMR DTC experiment of Refs.~\cite{rovny_observation_2018, RovnyPRB}. A standard NMR setup entails nuclear spins $\vec{I}_i$ located on sites $i$ of a crystalline lattice, interacting via dipolar interations, $J_{ij} \sim \frac{\mu_0\gamma_i \gamma_j}{4\pi |\vec{r}_{ij}|^3} \left(\vec{I}_i \cdot \vec{I}_j -3 (\vec{I}_i \cdot \hat{\vec{r}_{ij}})(\vec{I}_j \cdot \hat{\vec{r}_{ij}})\right)$ for spins separated by the lattice vector $\vec{r}_{ij}$, where $\mu_0$ is the vacuum permeability, and $\gamma_i$ and $\gamma_j$ are the nuclear gyromagnetic ratios of the two
spins. In the NMR DTC experiment~\cite{rovny_observation_2018}, the spins are furnished by spin-1/2 ${}^{31}P$ nuclei in ammonium dihydrogen phosphate, and are arranged in a three dimensional crystalline lattice. As is typical of NMR experiments, the setup is subject to a strong magnetic field oriented along the $z$ direction (by convention). The Zeeman splitting of the nuclear spins from this applied field is several orders of magnitude larger than the strength of the dipolar interactions, and the Zeeman field leads to a very fast precession of all transverse components of the nuclear spins. Then, in the rotating frame of this large applied field, one can define a so-called `secular' Hamiltonian which takes a `XXZ' form for interactions between spins of the same type~\cite{rovny_observation_2018}: 
\begin{align}
    H_{\rm secular} &=  \sum_{i,j} \frac{\mu_0\gamma_i \gamma_j}{4\pi |\vec{r}_{ij}|^3}\frac{1}{2}[3 \cos^2(\theta_{ij})-1] \left(\vec{I}_i \cdot \vec{I}_j -3 I^z_i I_j^z \right)\nonumber\\
    &+  h \sum_i I_i^z   +\cdots,
\end{align}
where $\theta_{ij}$ is the angle between the internuclear vector $\vec{r}_{ij}$ and the $z$ axis (defined by the static external field), $I_i^z$ refers to the $z$ component of the spin $\vec{I}_i$ and $\cdots$ refer to other couplings between other types of nuclear spins and higher order terms. Note that the secular Hamiltonian conserves the total $z$ component of the magnetization, $M = \sum_i I_i^z$. 

The DTC experiment~\cite{rovny_observation_2018, RovnyPRB} prepares a weakly magnetized mixed initial state at high temperature. The spins interact via $H_{\rm secular}$ for a time-period $T$ and are then periodically subject to a near perfect $\pi$-pulse that globally flips all spins, with a deviation $\epsilon$. Despite the systematic deviation in the rotation angle, the experiment observes a period doubled signal locked at a frequency $\omega/2$ for strong enough interactions $J$, one characteristic signature of time-crystalline order~\cite{briefhistory}. For weaker interactions, the system crosses over to a regime with ``beating" at a frequency that tracks $\epsilon$ instead of robust period doubling---thereby crossing over from a regime with a time-crystalline signature to one without. 

For concreteness, we will work with a tractable minimal model inspired by the NMR drive in which the essential ingredients for producing DTC behavior via $U(1)$ prethermalization are manifest.  
We study a driven one dimensional system of spin degrees of freedom on sites $i$, $S_i^\alpha = \frac{1}{2}\sigma_i^\alpha$ where $\alpha = x,y,z$ and $\sigma^\alpha$ are Pauli spin 1/2 matrices.  The drive consists of three elements. The first are XXZ type nearest and (integrability-breaking) next-nearest neighbour interactions of respective strengths $J,J'$; the second a uniform magnetic field in the $z$ direction, $h \Sztot$; and the third a periodically applied global spin rotation by an angle $\theta$ about the $x$ axis, generated by $P^x_\theta$. Our choice to work in one dimension with truncated-range interactions is for numerical tractability; our conceptual framework is equally applicable to higher dimension and longer-range interactions. 

The model drive is: 
\begin{align}  
    P^x_\theta&= e^{-i \theta \Sxtot},\nonumber \\
    H_{c} &= J\sum_{i=1}^{L-1}(\X_i \X_{i+1} + \Y_i \Y_{i+1} -2 \Z_i \Z_{i+1}) \nonumber\\
    &+J'\sum_{i=1}^{L-2}(\X_i \X_{i+2} + \Y_i \Y_{i+2} -2 \Z_i \Z_{i+2}),
    \label{eq:PHc}
\end{align}
with $J =1, J' = 0.5$. The resulting Floquet unitary, which is the stroboscopic time evolution operator over one period, is given by 
\begin{align}
    U(T) &= P^x_\theta e^{-i T_1 (H_{c} + h \Sztot)} \nonumber \\
 &= P^x_\theta e^{-i h T_1 \Sztot} e^{-iT_1H_{c}}\, 
    \label{eq:Floquetdrive}
\end{align}
where the second line  follows from the first because $[H_c, \Sztot] = 0$ justifies the separation of exponentials. 
In what follows, we vary the period $T_1$ and  field $h$, while considering small, $\theta =0+\epsilon$, and nearly maximal, $\theta = \pi+\epsilon$ (``$\pi$ pulse''), spin rotation angles. These are detuned by a small amount $\epsilon=0.1$, unless otherwise stated, to address the stability of the phenomena we discuss. 
The exact $\pi$-pulse, $P^x_\pi \sim \prod_i \sigma_i^x$, enacts a perfect flip of all spins in the $z$ basis.  For $\theta = \pi+ \epsilon$, the flip has a systematic deviation, as in NMR experiment. 

Let us discuss some salient features of this drive. 

First, for $\epsilon=0$, the evolution can be identified with that of a static Hamiltonian with perfect U(1) symmetry. This is trivially true when $\theta = 0$, in which case the problem reduces to an undriven one, $U(T) = e^{-i T_1 (H_c+hS^z_{\rm tot})}$. For the ``flipped" case with spin rotation angle $\theta=\pi$, this is still true if the two period unitary is considered, $U(T)^2 = e^{-i 2 T_1 H_c}$, which follows from the fact that $[H_c, P^x_\pi]=0$ and $\{S^z_{\rm tot}, P^x_\pi\}=0$. In the latter case, the U(1) symmetry can be used to achieve perfect period doubling (or ``time-crystalline'') dynamics with the global magnetization flipping every period: $M(nT)=  (-1)^n M(0)$, where $M = \sum_i \sigma_i^z$, $n$ is an integer, and $M(nT)  =  U^\dagger(nT) M U(nT)$. Again, this follows simply because $U(2T)$ commutes with $M$ while $U(T)$ anticommutes with $M$ due to the action of the $\pi$-flip, and does not rely on symmetry breaking. 

On the other hand, for any non-zero $\epsilon$, the system is genuinely driven and will eventually approach the infinite temperature ensemble, $\rho \propto \mathbb{I}$. 
\emph{The challenge is thus to generate long timescales, $t_h$ and $t_m$, both for the approximate non-conservation of energy and the approximate non-conservation of $\Sztot$ respectively. }

Second, note that the second line of Eq.~\eqref{eq:Floquetdrive} implies that $hT_1$ is only defined modulo $2\pi$, and hence cannot be made arbitrarily large for this drive: \emph{there is no simple high-field limit}.  
We will show that the dynamics in this system can be explained via an approximate long-lived conservation of $\Sztot$. 
However, the approximate conservation of $\Sztot$ -- the central emergent feature -- is not due to a large field $h$. Instead, it is the smallness of the detuning $\epsilon$ of the global spin rotation, which controls \emph{both} the strength of driving \emph{and} the strength of the $\Sztot$ violation. We emphasize that while a large magnetic field is used in the NMR set up to obtain an interaction Hamiltonian $H_c$ that conserves $\Sztot$ within the secular approximation,  the periodic drive as a whole comprises both the interaction Hamiltonian (with the field) and the global spin rotation --- and there is no simple large field limit to obtain $\Sztot$ conservation for $U(T)$ as a whole.

Third, in the low-frequency regime when $\omega =2\pi/T_1 \ll \{J,h\}$, the experiment finds that the envelope of $M(nT)$ decays {exponentially} in time so there is no long-lived signal. In this regime, there is no quasi-conservation of energy and
$|M(t)|\sim e^{-t/t_d}$, with a decay time $t_d \sim 1/\epsilon^2$. This is the ``dephasing regime"~\cite{Misha3T}. The basic picture is that the operator $\Sztot$ gets ``rotated away" from the $z$ axis by an angle $\epsilon$ every drive cycle under the action of $P^x_\epsilon$. The component of the polarization in the $XY$ plane gets dephased under the action of $H_c$, while the component parallel to $z$ remains conserved while $H_c$ acts. This gives a decay $M(n)\sim \cos(\epsilon)^n$, consistent with the observed exponential decay upon expanding in small $\epsilon$. In the next section, we will work instead in the high-frequency limit which allows us to enter a prethermal regime.

\begin{figure*}
    \includegraphics[width=\textwidth]{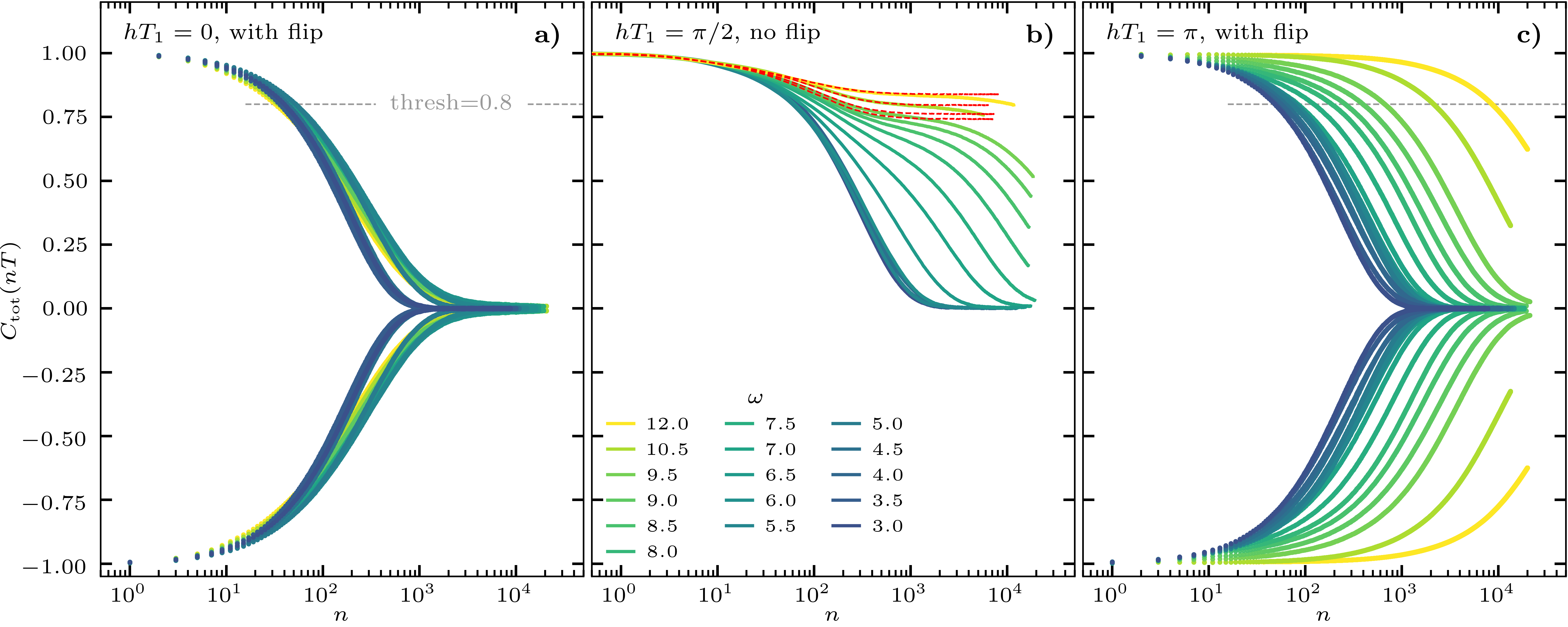}
	\caption{Survival of the total magnetization $C_{\rm tot}(nT)$ (defined in Eq.~\eqref{eq:Ctot}), stroboscopically observed, in a chain of length $L=20$ under the NMR Floquet drive for different driving frequencies $\omega=2\pi/T_1$ \eqref{eq:Floquetdrive}. The detuning of the spin rotation is $\epsilon=0.1$. 
    \textbf{a)} Without a field ($h=0$) and with an approximate global flip $P^x_\theta$, $\theta={\pi+\epsilon}$, corresponding to the presumptive parameters of the NMR experiment, the magnetization dies off quickly with little dependence on the driving frequency. 
	\textbf{b)} At half the maximum field, $h=\pi/(2T_1)$, and with $\theta=0+\epsilon$, the survival of the magnetization is enhanced. Dashed lines show the evolution with the time averaged leading-order effective Hamiltonian. 
\textbf{c)} With the maximal field $h=\pi/T_1$ and with an approximate global spin flip $\theta=\pi+\epsilon$, we observe a prethermal time crystalline signal with dramatically \emph{enhanced lifetime}, by more than $100\times$ compared to (a). The lifetime shows an exponential dependence on driving frequency (Fig.~\ref{fig:nmr2}), a hallmark of prethermalization. 
\label{fig:nmr1}}
\end{figure*}

\begin{figure}
    \includegraphics[width=\columnwidth]{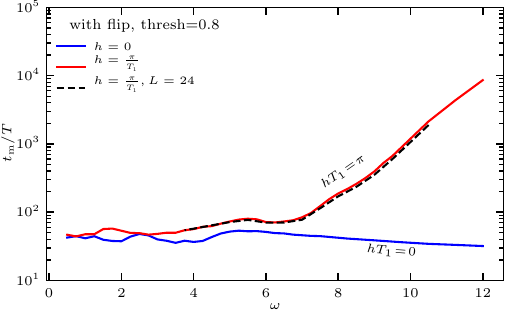}
	\caption{ Number of driving periods needed to reach a magnetization threshold of $C_{\rm tot} = 0.8$, which serves as an estimate for $t_m$ for the data in Fig.~\ref{fig:nmr1} with no ($h=0$) and maximal field ($h=\pi/T_1$). At large frequencies, the maximal field data shows a greatly increased $t_m$ with an exponential dependence on $\omega$. The dashed curves show additional data for longer chains ($L=24$). 
\label{fig:nmr2}}
\end{figure}

\section{Regimes of Thermalization}
\label{sec:heating}

In this section, we discuss various regimes of thermalization for the model drive in Eq.~\eqref{eq:Floquetdrive} in the high-frequency regime $\omega \gg \{J,h\}$. The demonstration of a long-lived U(1)-stabilised DTC signal proceeds in several steps. Our starting point is an analysis of the NMR experiment with its period-doubled response (Case 1, Fig~\ref{fig:nmr1}(a)). We then demonstrate how to extend its lifetime dramatically into a bona-fide prethermal signal by adjusting the drive parameters. This proceeds in two steps.  We first arrange for the appearance of a prethermal regime by adjusting the drive (Case 2, Fig~\ref{fig:nmr1}(b)) to yield an effective Hamiltonian with emergent approximate U(1) conservation, but no period doubling. The second step (Case 3,  Fig~\ref{fig:nmr1}(c)) recreates the period doubled DTC response, now with a parametrically longer lifetime, by adjusting the applied Zeeman field. The resulting dramatic increase in lifetime is a consequence of the effective Hamiltonian, at leading order, being perfectly U(1) symmetric. 

As discussed previously, the high frequency regime $\omega \gg \{J,h, \theta\}$ allows one to define a quasilocal time-independent effective Hamiltonian associated with the quasiconservation of energy~\cite{DimaPrethermal_rigorous,DimaPrethermal_Heff,MoriPrethermal1, MoriPrethermal2}. The dynamics are well approximated by $H_{\rm eff}$ up to the `heating time' $t_h \sim e^{\omega/J}$. We note that the existence of this prethermal regime requires $\omega$ to be much bigger than all local energy scales in the time-dependent Hamiltonian Eq~\eqref{eq:PHc}. However, this is not strictly true when $\theta= \pi + \epsilon $, which is the regime in which one gets a non-trivial DTC phase. This issue is also encountered when considering the existence of Floquet MBL in the DTC phase, and the resolution is that one needs to eliminate the large frequency effect of the $\pi$-pulse first, which is conveniently done by either working in a ``toggling" frame which rotates by $P^x_\pi$ each time a global rotation is applied or, equivalently, by considering the time evolution over two periods:
\begin{align}
    U(2T) &= P^x_\pi P^x_\epsilon e^{-i h T_1 \Sztot} e^{-iT_1H_{c}} P^x_\pi P^x_\epsilon e^{-i h T_1 \Sztot} e^{-iT_1H_{c}}\nonumber\\
    &=  P^x_\epsilon e^{+i h T_1 \Sztot} e^{-iT_1H_{c}} P^x_\epsilon e^{-i h T_1 \Sztot} e^{-iT_1H_{c}}
\end{align}
where we have used $[H_c, P^x_\pi]=0$ and $\{\Sztot, P^x_\pi\}=0$ to eliminate the large $\theta$ field. Then, one can define an effective Hamiltonian which, to leading order in $1/\omega$, is a time-average: 
\begin{align}
    \Heff &= \Ho + O(1/\omega) \nonumber\\
    \Ho &\propto  \frac{T_1}{T_1+\epsilon} \left[H_c + 
 \frac{\epsilon_{\rm eff}}{T_1} \sum_i S^x_i + h_{\rm eff} \sum_i \Z_i\right] + O(1/\omega).
    \label{eq:Heff}
\end{align}
 Note that $\Heff$ is a generic thermalizing Hamiltonian with no disorder and no MBL. Higher order terms in $\Heff$  make it quasi-local with a decaying range of interactions. 
For the ``unflipped" case when $\theta =\epsilon$, we consider a single period $U(T)$ \eqref{eq:Floquetdrive} and the proportionality constant in $\Ho$ is 1, and $h_{\rm eff} = h$ and $\epsilon_{\rm eff} = \epsilon$.  
On the other hand, when $\theta =\pi + \epsilon$, we define $\Heff$ with respect to $U(2T)$ the proportionality constant for $\Ho$ is 2, and the values $h_{\rm eff}$ and $\epsilon_{\rm eff}$ depend on $h$ and $\epsilon$ as discussed below.

With this in hand, we can predict drives for which prethermal -- including time-crystalline --  phenomena occur, and also understand the status of the experimental protocol in this regard. One of our main messages is that if $\Heff$ shows an approximate long-lived $U(1)$ conservation, then $M(t)$ will saturate at a non-zero constant value on some (typically short) time-scale under evolution with $\Heff$, for {\it all} initial states that start with a non-zero magnetization density. In the flipped case where $\Heff$ captures the dynamics over two periods (and hence over either even or odd times), a non-zero thermal value for $M(2nT)$ implies period doubled oscillations when considering both even and odd times because $P^x_\pi M P^x_\pi = -M$. 

We will denote by $t_m$ the time-scale on which the approximate $U(1)$ conservation is destroyed. This can arise via two mechanisms: (i) the system heats to the infinite temperature ensemble with no conservation laws, and $\Heff$ ceases to be a good description, which occurs on time-scale $t_h$. Expectation values for all observables, including $M(t)$ become trivial at this time;  (ii) the system thermalizes to the `true' equilibrium thermal ensemble for $\Heff$ which does not conserve $U(1)$ symmetry (higher-order terms in $\Heff$ are not fine-tuned in any way, and will generally break the symmetry). At this time, denoted $t_{\rm th}$, $M(t)$ will decay to zero for all initial states at infinite temperature with respect to $\Heff$, including those that started with a finite magnetization density. Note that $t_{\rm th}$ will be set by a combination of $\epsilon$ and $\omega$ and could be made larger than $t_h$ for small enough $\epsilon$. Thus, $t_m \sim \min[t_{\rm th}, t_h]$ will be set by the faster of the two processes above, and this sets the lifetime of the DTC response for the period-doubled case. 

To examine the presence of $U(1)$ conservation in an initial state independent manner, we consider the normalized quantity~\cite{DimaPrethermal_rigorous}
\begin{align}
 \Delta(nT) &\equiv  \frac{1}{2L} ||M(nT) - M(0)|| \nonumber \\
 &=   \frac{1}{2L} \frac{1}{2^L} \mbox{Tr} [(M(nT)-M(0))^{\dagger}(M(nT) - M(0))]\nonumber \\
 &= 1 - \frac{1}{L} \frac{1}{2^L} \mbox{Tr} [M(nT)M(0)] \nonumber \\
 & \equiv 1- C_{\rm tot}(nT).
 \label{eq:Ctot}
 \end{align}
Here $||\; ||$ denotes the Hilbert-Schmidt operator norm, and $\Delta(nT) = 0$ when $M$ is strictly conserved in which case $M(t)=M(0)$. The third line uses the fact that $\mbox{Tr}[M^\dagger(t) M(t)] = \mbox{Tr}[M(0)M(0)] =  \sum_{ij} \mbox{Tr}[\sigma_i^z \sigma_j^z]= L2^L$ by the cyclicity of trace and the tracelessness of the Pauli operators. This expression relates the conservation of global magnetization to the infinite temperature autocorrelator of $M(t)$, which is experimentally measurable. 
We note that, more precisely, an approximate $U(1)$ conservation will manifest itself as a dressed quasi-local operator $\tilde{M}$ that is conserved for a long-time $t_m$, and $\tilde{M}$ only agrees with $M$ to leading order in a small parameter~\cite{DimaPrethermal_rigorous}. 

We  study the normalized autocorrelator $C_{\rm tot}(t)$ defined above, and the deviation of the (absolute value) of this quantity from 1 is a proxy for the non-conservation of $M$ in the system. For an efficient numerical simulation of the system, we use quantum typicality~\cite{bartsch_dynamical_2009,reimann_dynamical_2018,luitz_ergodic_2017,luitz_absence_2017} to replace the trace so that
\begin{equation}
    C_{\rm tot}(nT) 
    \approx \frac{1}{L} \braket{\tilde \psi|M(nT) M |\tilde \psi},
    \label{eq:typ}
\end{equation}
where $\ket{\tilde\psi}$ is a random (Haar measure) state, typical for infinite temperature. We can then efficiently simulate the dynamics using numerically exact Krylov space time evolution technique\cite{nauts_new_1983,moler_nineteen_2003,luitz_ergodic_2017} to calculate the action of matrix exponentials on wave functions. This allows us to access large systems of sizes $L=20-24$, beyond those accessible to ED. Accessing these large sizes is particularly crucial in numerical studies of prethermalization which require us to work in the regime $J \ll \omega \ll JL$, where the first inequality is required to get a long-time scale $t_h$, and the second is required to keep the $O(1)$ frequency smaller than the extensive many-body bandwidth so as work in a sensible thermodynamic limit. In practice, the MB bandwidth is a factor of 5-10 larger than the frequency for the sizes we can achieve.  
The typicality approximation in Eq. \eqref{eq:typ} is exponentially accurate in terms of $L$, and for our Hilbert space dimensions $>10^6$, using a single wavefunction $\ket{\tilde\psi}$ is sufficient.
We now examine the behavior of $C_{\rm tot}(nT)$ for several different cases.

\subsection{Case 1: $hT_1 =0$, Experimental Choice} Let us start with the choice of parameters in the NMR experiment, in which a large Zeeman field is applied but the field is tuned so that $hT_1=0 \; \mbox{mod} \; 2\pi$. This corresponds to $h_{\rm eff}=0$ and $\epsilon_{\rm eff} = \epsilon$ in $\Ho$, Eq.~\eqref{eq:Heff}. Note that one also obtains $h_{\rm eff}=0$ if there is a $\pi$ pulse present, because the $z$ field flips sign and gets ``echoed" out to leading order under the action of the $\pi$-flip. 
However, $h_{\rm eff}=0$ is maximally non-ideal from the point of view of the ``usual" mechanism for generating an approximate $U(1)$ conservation in a time-independent Hamiltonian, which attempts to engineer a large separation of scales between different $\Sztot$ sectors by subjecting the system a large magnetic field~\cite{DimaPrethermal_rigorous}. Indeed, the lifetime of TC response seen is the shortest for this case (Fig.~\ref{fig:nmr1}(a)). 

To wit, consider starting from an initial state that is at near infinite temperature with respect to $\Heff$, but has a net magnetization density, similar to the experiment. Then, even if there is a long-lived $\Heff$ with $t_{h} \sim \exp(\omega/J)$, the appropriate thermal value for $M$ with respect to $\Heff$ is 0, and $M(2nT)$ will thermalize to zero on a time-scale, $t_{\rm th}$, set by the destruction of $\Sztot$ conservation in $\Heff$. In the absence of a large scale $h_{\rm eff}$,  this time scales as $t_m \sim t_{\rm th} \sim 1/\epsilon^2$ and depends \emph{polynomially} on $1/\epsilon$ by standard Golden-Rule type reasoning. 

In other words, even though $t_h$ scales exponentially with $\omega$, we do not expect $t_m \sim t_{\rm th}$ to show a strong $\omega$ dependence. The $\omega$ independence is borne out by the numerical data in Figs.~\ref{fig:nmr1}a and \ref{fig:nmr2}) and explains the relatively short lifetime of the time crystalline response due to the absence of any exponential scaling in $t_m$. We note that since the actual experiment only measures only 100 cycles, it still sees a finite Fourier peak at $\pi$ corresponding to the transient period doubling. 
Of course, in principle, one could also consider a regime with small enough $\epsilon$ such that $t_{\rm th} > t_{h}$. In this regime, the conservation of magnetization is destroyed due to $\Heff$ ceasing to be a good description rather than the destruction of $\Sztot$ conservation within $\Heff$ \emph{i.e.} $t_m\sim \min(t_h, t_{\rm th})$; this corresponds to a window with a prethermal exponential dependence, $t_m \sim t_h  \sim \exp(\omega)$ for the smallest range of $\epsilon$'s. 

A prethermal regime can be enhanced, however, through modifying the magnetic field $h$. 

\subsection{Case 2: $hT_1 =\pi/2$, Generation of a prethermal regime} 
Let us begin with the discussion of the nontrivial half-maximal value (due to the compactness of the unit circle) $hT_1 = \pi/2$. Here, we consider the case without a $\pi$ pulse, i.e. we include only a small spin rotation by $\theta=\epsilon=0.1$ per period $T$ of the drive. This prevents the field from being ``echoed out", so that $h_{\rm eff} = h = \pi/2T_1$. Adding this field separates the different $S^z_{\rm tot}$ sectors and makes the spin flip terms of strength $\epsilon$ more off-resonant and hence more ineffectual at destroying $S^z_{\rm tot}$ conservation.  However, as mentioned earlier, this field cannot be made parametrically large as is required for a bona-fide prethermal $U(1)$ regime in $\Heff$ (\emph{i.e.} one with an exponentially long-lived in $h$ $U(1)$ conservation). Instead, it still the case that $\Sztot$ conservation in $\Heff$ is destroyed on a polynomial in $1/\epsilon$ time-scale. 

However, for this case, we show that prethermalization in the dynamics of $M(t)$ can be achieved by directly coupling to prethermalization of energy and relying on a notion of temperature, rather than relying on $U(1)$ conservation. We note that the temperature can be quite high in this case, much higher than ordering temperatures for spontaneous symmetry breaking (in higher dimensions). 

Consider again an initial state at a finite magnetization density. Now, due to the presence of the $z$ field in $\Heff$, this state is \emph{also} at a finite energy density and hence temperature. Thus, $M(nT)$ will show an approximate ``plateau" to a \emph{non-zero} thermal value appropriate to the temperature of the initial state, before relaxing to zero at a later time scale $t_m \sim t_h \sim \exp[\omega/J]$, at which $\Heff$ ceases to be a good description. 

Indeed, Fig.~\ref{fig:nmr1}b) shows an initial relaxation of $C_{\rm tot}(nT)$ to a plateau at short times, followed by a later decay to zero as expected for an infinite temperature state with no additional conservation laws. We have verified that this later-time scale scales as $t_m \sim \exp[\omega/J]$ (not shown). The dashed lines in the figure correspond to the thermal expectation value obtained by direct evolution with the time averaged Hamiltonian $H_\text{eff}$, and these match the plateau values as expected.

\subsection{Case 3: $hT_1 =\pi$, Prethermalization without temperature} Finally, we consider the maximal possible field (again due to the compactness of the unit circle) of $hT_1=\pi$, which  leads us to demonstration of prethermalization \emph{without} temperature, relying solely on $U(1)$ conservation. We also return to $\pi$-flip case to obtain a prethermal DTC. It turns out that for this value of $h$, the field segment of the drive \emph{also} realizes a $\pi$-pulse, but now in the $z$ direction: $e^{-hT_1\Sztot} \sim P^z_\pi = \prod_i \sigma_i^z$. 
Again, let us look over two periods. 
Crucially, the effect of the $z$ $\pi$-pulse is to ``echo-out" the $\epsilon\Sxtot$ term. Thus, $\Ho$ over two periods has both $h_{\rm eff}=0$ \emph{and} $\epsilon_{\rm eff} = 0$ so that the (leading order) $\Heff$ exactly conserves $\Sztot$. 

Now, if one prepares initial states at infinite temperature with respect to $\Heff$ but finite magnetization density (which is possible because $h_{\rm eff} = 0$),  then the initial value of $M$ will persist for the time that the dynamics is approximately governed by $\Heff$. Then, at time $t_m \sim t_h\sim \exp(\omega/J)$, the magnetization decays to zero once $\Heff$ ceases to be a good description. As explained earlier, if one looks over both even and odd times, then oscillations are visible. This is confirmed in Fig.~\ref{fig:nmr1}c).

\begin{figure}
    \centering
    \includegraphics[width=\columnwidth]{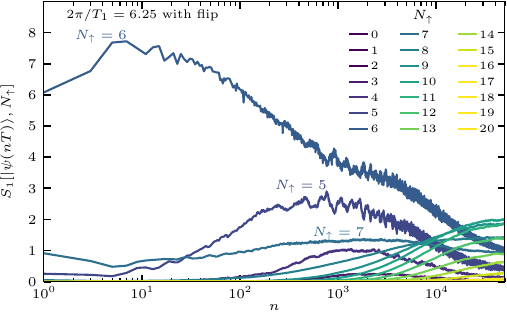}
    \caption{Time dependence of the sector (labelled by the number of up spins $N_\uparrow$) resolved participation entropy $S_1\left[ \ket{\psi(nT)}, N_\uparrow \right]$ of the wavefunction $\ket{\psi(nT)}$ starting from the initial state $\ket{00100100100100100100}$ (i.e.\ in the sector with $N_\uparrow=6$) under the NMR Floquet drive with a frequency of $\omega=6.25$, $hT_1=\pi$, and an approximate global spin flip after each period $\theta=\pi+\epsilon$ with $\epsilon=0.18$. The wavefunction spreads quickly within one sector, before slowly spreading over several sectors. 
    \label{fig:nmr-participation} }
\end{figure}

Put differently, when $\Heff$ has $U(1)$ conservation, the equilibrium ensemble of $\Heff$ is characterized by both a temperature $\beta^{-1}$ and a chemical potential $\mu$. One can prepare initial states that have $\beta = 0$, but have finite $\mu \neq 0$, and hence can show a persistent magnetization --- thereby separating the notion of prethermalization from temperature by allowing for a separate thermodynamic parameter. 

As is already visible by direct inspection of the time traces of $C_{\rm tot} (nT)$ in Fig. \ref{fig:nmr1} a) and c), the lifetime of the approximate conservation of $\Sztot$ and consequently the time crystalline behavior is strongly enhanced by the presence of a magnetic field in $z$ direction. \emph{In other words, a small modification of the applied field in the NMR experiment can lead to an exponentially greater DTC lifetime!~\footnote{We note, however, that experimentally seeing this large enhancement might be challenging. Even with $\epsilon=0$ which should realize an `ideal' period doubled signature forever, a finite duration for the $\pi$ pulses during which the interaction Hamiltonian is still present leads to a decaying signal in the actual experiment. }}

Fig. \ref{fig:nmr2} shows a direct comparison of the TC lifetimes in the case of zero and maximal field by extracting the time it takes for $C_{\rm tot}(nT)$ to decay to a threshold value of $0.8$. At high driving frequencies $\omega\gtrsim 6$, we observe an {\it exponential} scaling of the lifetime with the frequency in the presence of the field -- the characteristic signal of prethermalization --  while without a field there is only a weak frequency dependence. The colored lines are extracted from the data for a chain of length $L=20$ in the other panels of \ref{fig:nmr1}, while the black dashed line stems from the analysis of a larger system of size $L=24$. Note the negligibly small system size dependence, which is to be expected as prethermalization is sensitive to the ratio of $O(1)$ parameter sizes rather than the system size. 

Finally, we note that the exact conservation of $\Sztot$ in $\Heff$ when $hT_1=\pi$ is only true to leading order in $1/\omega$. Higher order corrections at $O(\epsilon J/\omega)$ will again cause $\Sztot$ to be destroyed within $\Heff$ on some polynomial time-scale $t_{\rm th} \sim \omega^2/\epsilon^2$. For very small $\epsilon$'s and large $\omega$'s such that $t_{\rm th} < t_h$, we will find that $t_m \sim t_{\rm th}$ does not show an exponential dependence on $\omega$. However, because the destruction of $\Sztot$ conservation on time-scales $t_{\rm th}$ only occurs due to higher order corrections in $\Heff$, in practice one can still isolate a large prethermal window where $t_m \sim t_h \sim \exp(\omega)$, as is visible from Fig.~\ref{fig:nmr1} c). In the limit that $\epsilon \rightarrow 0$, this window in $\omega$ can be made arbitrarily large. 

Although discussed in the context of our model drive, the mechanism outlined above is very general. One can consider large families of drives for which the leading terms in $\Heff$ have a desired symmetry, with corrections only coming in at higher orders in $1/\omega$. When the strength of these corrections is further controlled by a small parameter $\epsilon$, it is possible to tease out an exponentially large window in $\omega$ for small enough $\epsilon$ during which the symmetry is approximately conserved. 

We next turn to a more detailed verification of our picture, and its stability.
First, Fig.~\ref{fig:nmr-participation} provides visually compelling direct evidence of the prethermal mechanism involving approximate U(1) conservation. It displays the participation entropy in the computational $z$ basis $\{\ket{i}\}$ upon starting from a specific basis state. The participation entropy of a wavefunction $|\psi\rangle$  in each magnetisation sector $\mathcal{H}_{N_\uparrow}$ is defined as $S_1[N_\uparrow] = -\sum_{i : \ket{i} \in \mathcal{H}_{N_\uparrow}} |\braket{i|\psi}|^2 \ln |\braket{i|\psi}|^2$, where $\mathcal{H}_{N_\uparrow}$ containing all basis states with $N_\uparrow$ spins pointing up. We find two distinct timescales: the wavefunction {\it very quickly} delocalises in its initial magnetisation sector, and then, much more slowly, leaks into increasingly distant other sectors, which is clearly visible in the delayed onset of the growth of the participation entropy for different magnetization sectors.

\begin{figure}
\includegraphics[width=\columnwidth]{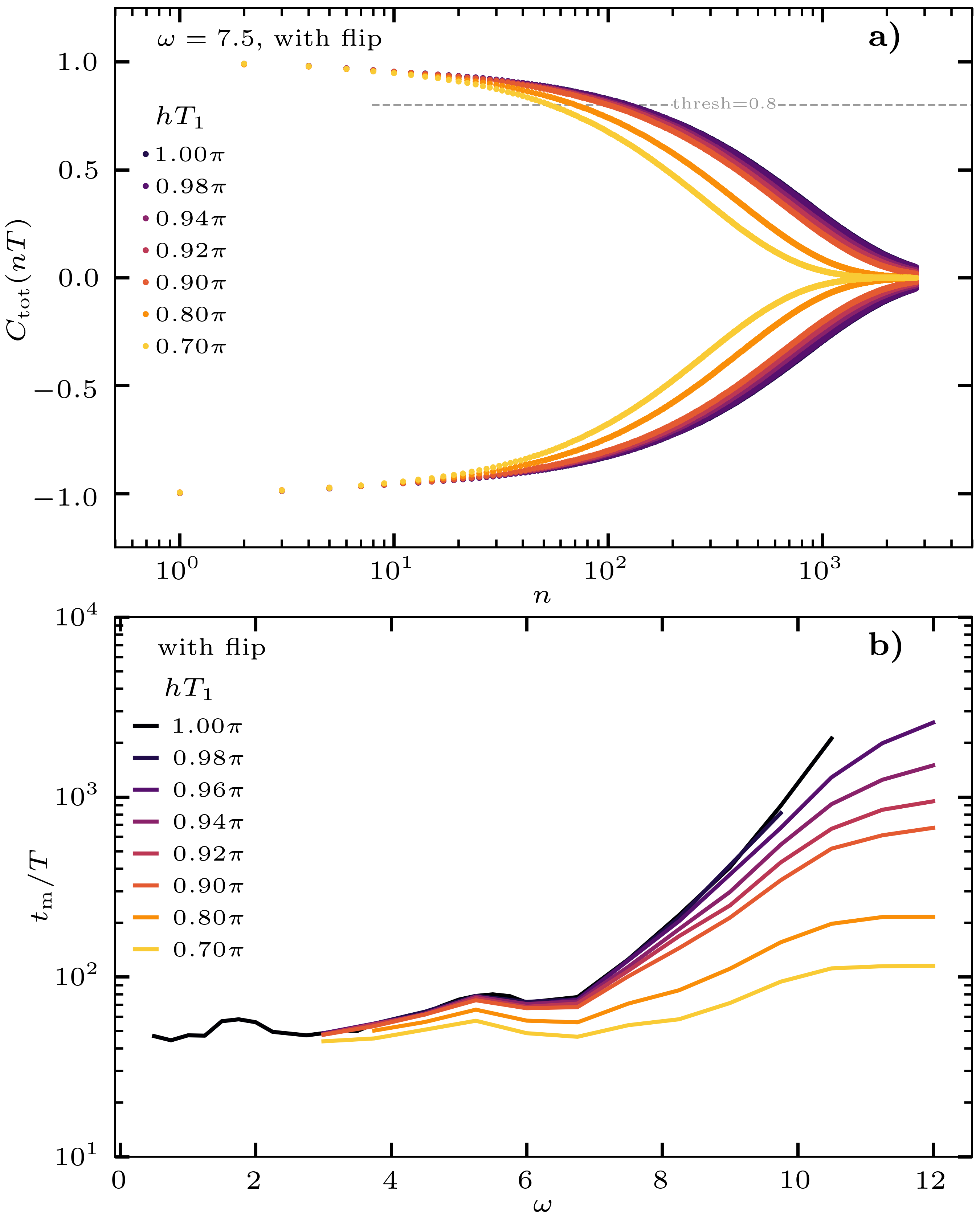}
\caption{Robustness of prethermal regime to drive imperfections. 
\textbf{a)} Survival of the total magnetization $C_{\rm tot}(nT)$ in a chain of length $L=20$ under the NMR Floquet drive \eqref{eq:Floquetdrive} for driving frequency $2\pi/T_1=7.5$ and magnetic fields $h$ detuned from the optimal field $h = \pi/T_1$. The detuning of the spin rotation $P^x_{\pi+\epsilon}$ is $\epsilon=0.1$.
	\textbf{b)} Number of driving periods needed to reach a thresholds of 0.8 for the same data as in panel a). 
\label{fig:nmr-detuning}	}
\end{figure}

Finally, we note that a judiciously chosen $z$ field to realize a $U(1)$ conservation in $\Heff$ is both experimentally feasible, and reminiscent of various ``dynamical decoupling" schemes for Hamiltonian engineering. However, our results are not limited to a fine-tuned $z$ field. Fig.~\ref{fig:nmr-detuning} shows the stability of the prethermal DTC lifetimes to detuning from the maximal field. While for maximal field the lifetimes are optimal as expected, the exponential scaling of the lifetimes with $\omega$ is still visible down to a detuning of about 10 percent from the maximal field.

\section{Distinguishing between prethermal and MBL TCs}
\label{sec:distinguish}
For realistic experiments with a lifetime limited by extrinsic factors, it may often be the case that prethermal time window is longer than the experimental lifetime. Thus, the question naturally arises on how to distinguish a prethermal DTC from a bona fide (MBL-localised) infinitely long-lived one, and also how to distinguish between prethermal $U(1)$ DTCs (this work) and prethermal SSB DTCs relying on spontaneous symmetry breaking (Ref.~\cite{else_prethermal_2017}). Indeed, all three DTC experiments thus far (on trapped ions, diamond NV centers and NMR spins) nominally observe very similar experimental signatures, but for apparently different reasons.  

To achieve this goal, we avail ourselves of the fundamentally distinct origin of the respective longevities. While the emergence of locally conserved quantities -- the l-bits -- underpin MBL~\cite{Huse14, Serbyn13cons, Imbrie2016}, the prethermal $U(1)$ DTC only offers a global conservation law which is not in conflict with local spin diffusion. The prethermal SSB DTC relies on yet a distinct mechanism which requires low-temperature initial states and spontaneous symmetry breaking. 

\begin{figure}
    \includegraphics[width=\columnwidth]{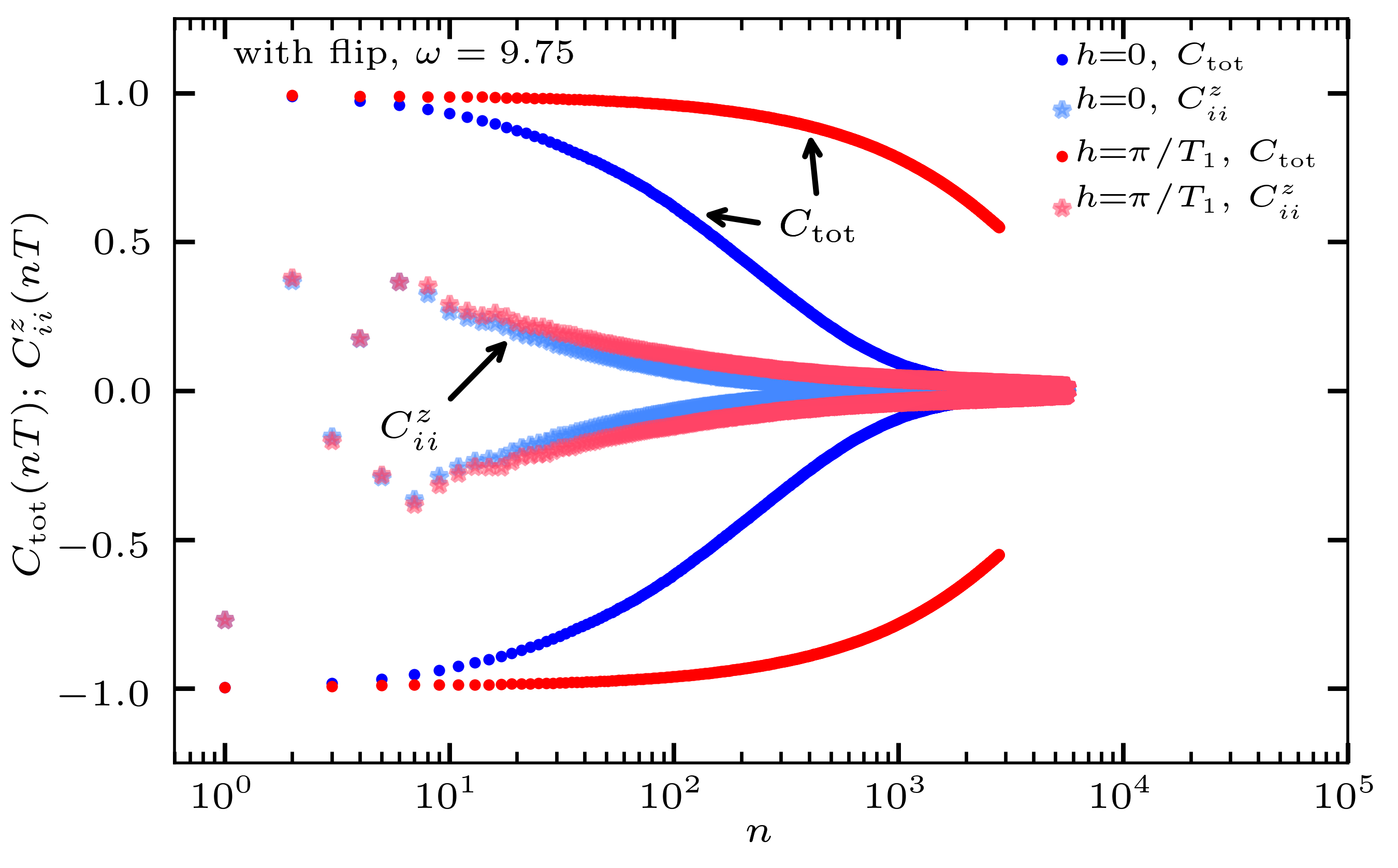}
	\caption{
Comparison between the survival of the global magnetization $C_{\rm tot}(nT)$ and the local correlation function  $C^z_{ii}=\frac{1}{2^L}\Tr\left( Z_{i}(nT) Z_i\right)$. The former can decay much more slowly when there is an approximate global $U(1)$ conservation, while the latter decays swiftly due to fast thermalization \emph{within} $U(1)$ sectors as shown in Fig.~\ref{fig:nmr-participation}. This is in contrast to a many-body localized time crystal where both local and global autocorrelators oscillate with a finite amplitude even at infinitely late times. 
    \label{fig:nmr-ztot}}
\end{figure}

The crispest way to distinguish between these mechanisms is by considering a variety of different initial states and measuring \emph{local} spin autocorrelators in the $z$ basis. If we start with a random infinite temperature product state of $z$ spins in the $\Sztot=0$ sector, only an MBL DTC will display oscillations in local autocorrelators $\langle \sigma_i^z(nT) \sigma_i^z\rangle$. These states are too high in temperature for prethermal SSB DTCs, and they have $\Sztot = 0$ leading to zero net magnetization density for the $U(1)$ DTC. Fig.~\ref{fig:nmr-ztot} displays a simulation of both local and global infinite temperature spin autocorrelators in the $z$-basis for the NMR model. In all settings displayed, we find that the local version decays much more quickly than the approximately symmetry-protected global one, while such a decay is precluded for the MBL version.  

Finally, to distinguish between prethermal $U(1)$ TCs and prethermal SSB TCs, one should start with initial states with $\Sztot = 0$, but still at a low temperature with respect to $\Heff$ (say states with a single domain wall in the center of the chain). Local autocorrelators in the $U(1)$ TC will thermalize within the $\Sztot=0$ sector, showing no net magnetization and zero amplitude of oscillations. On the other hand, the prethermal SSB DTC will show oscillations in local correlators starting from such states, with occasional ``phase slips" at late times due to slow coarsening dynamics of domain walls. 

By contrast, if one starts from a polarized initial product state, then all three categories give virtually indistinguishable signatures. This is a drawback of existing TC experiments on disordered systems which only consider a very limited class of initial states. Indeed, the need for more fine-grained experimental diagnostics was made particularly apparent in a recent theoretical study of the trapped ion experiment~\cite{briefhistory}. The trapped ion setup tries to realize an MBL DTC phase by engineering a Floquet Ising drive with imperfect $\pi$ flips, and disorder in the \emph{longitudinal} $z$ fields. However, the disorder in the fields is echoed out under the $\pi$-flip to leading order, so that the model does not realize an MBL TC but rather looks to be a prethermal SSB DTC. If the experiment had been conducted for a wide variety of initial states (instead of only two low-temperature initial states) this difference would have been apparent. 

Finally, one can ask if the mechanism of $U(1)$ prethermalization might be at play in the trapped ion/NV center drives, since both models do have a $U(1)$ symmetry for $\epsilon=0$. Similar to the NMR experiment, the NV experiment can only measure a global polarizaration, and the experiment starts with a fully polarized initial state and observes an oscillating signal for $M(t)$ with a slowly decaying envelope. Indeed, an effective Hamiltonian for this model is also obtained by adding a large $z$ field and going to an appropriate rotating frame -- if this applied field were chosen to be $hT_1=\pi \mod 2\pi$ (which removes the deviation $\epsilon$ to leading order), then one would observe an enhancement of the DTC signal for this experiment as well. However, in the NV model, thermalization is a critically slow process due to disorder~\cite{CriticalTCPRL} --- so that local autocorrelators would \emph{also} decay slowly, unlike the NMR experiment where the decay of local correlators is fast. However, neither experiment has access to site resolved local autocorrelations, so this difference between their thermalization mechanisms cannot be experimentally verified. The trapped ion experiment \emph{does} measure local autocorrelators, but numerics for this model on different initial states are consistent with a prethermal SSB DTC rather than a prethermal $U(1)$ DTC~\cite{khemani_phase_2016}. In principle, the trapped ion experiment could be repeated with a variety of different initial states to elucidate this difference.

\section{Conclusions}
\label{sec:conclusion}
In summary, we have analyzed in detail a scenario relevant for the optimization of NMR experiments on prethermal discrete time crystals in periodically driven quantum many-body systems. We argue that at high enough driving frequency, an optimal magnetic field exists which stabilizes an approximate U(1) conservation law and bears the potential to enhance the lifetimes of time crystalline behavior by two orders of magnitude (Fig.~\ref{fig:nmr1}(a,c)). This optimization represents a small modification of the existing NMR experiment and should be achievable in practice. 

One of our main contributions is to connect previous NMR insights with more rigorous theories of prethermalization to (i) demonstrate a large parameter window with an exponential lifetime for an emergent conservation law, even without large magnetic fields in $\Heff$ (ii)  elucidate how this permits interesting prethermal dynamics, even at infinite temperature with respect to $\Heff$, and (iii) explain how these can be combined to obtain prethermal time crystals at high temperatures and without relying on the existence of symmetry breaking in $\Heff$. This, in turn opens up the possibility of realizing prethermal TCs in a much wider range of settings than is known thus far, because the need for SSB in thermalizing Hamiltonians comes with stringent constraints on allowed spatial dimensions and ranges of interactions due to Peierls-Mermin-Wagner type theorems. 

Finally, it is interesting to ask if the notion of an effective Hamiltonian can be dispensed with altogether to achieve drives with \emph{only} a $U(1)$ conservation without any notion of an effective Hamiltonian --- the most dramatic rendition of prethermalization without temperature. Indeed, consider a drive with an oscillating magnetic field of the form~\cite{AsmiStrongDriving}:
\begin{equation}
    U(T) = e^{-i \frac{T}{2} (H_c + \epsilon \Sxtot + h\Sztot)} e^{-i \frac{T}{2} (H_c + \epsilon \Sxtot - h\Sztot)}.
 \end{equation}
In this case, the strength of the field $h$ \emph{can} be made extremely large because of the presence of the non-commuting $\epsilon \Sxtot$ in both Hamiltonians. Now, if one works in the high-frequency limit, then the leading order $\Heff $ averages over both terms and reduces to the $h_{\rm eff} =0$ case considered earlier. On the other hand, in the \emph{low} frequency limit where such an averaging is not appropriate and $\Heff$ is not defined, each term of the drive can be made to conserve $U(1)$ in a crisp prethermal sense, thereby endowing $U(1)$ conservation to the drive as a whole. We have qualitatively verified numerically that this drive has enhanced $M(t)$ conservation at small rather than large frequencies. However, obtaining a quantitative agreement is limited by finite-size numerics due to our inability to access a regime where $J \ll h \ll \omega \ll JL $.

\section{Acknowledgements} We thank Sean Barrett, Robert Blum, Jared Rovny, Sarang Gopalakrishnan and Wen Wei Ho for discussions. RM thanks Arnab Das, Asmi Haldar and Diptiman Sen for collaboration on related topics. VK and SLS thank Curt von Keyserlingk and Matteo Ippoliti for collaboration on related topics. This work was in part supported by the DFG through  ct.qmat (EXC 2147, project-id 39085490). VK was supported in part by the Harvard Society of Fellows and the William F. Milton Fund. 
This research was also developed with funding from the Defense Advanced Research Projects Agency (DARPA) via the DRINQS program. The views, opinions and/or findings expressed are those of the authors and should not be interpreted as representing the official views or policies of the Department of Defense or the U.S. Government.
DJL thanks PRACE for awarding access to HLRS’s Hazel Hen computer based in Stuttgart, Germany under grant number 2016153659.
Our code is based on the PETSC and SLEPc libraries.

\bibliography{global,prethermal}

\end{document}